\documentclass[12pt,a4paper]{article}
\usepackage{geometry}
\include {epsf}
\def\b{\beta}

\def\D{\Delta}

\def\dmsq{\D m^2}

\def\Ar{\rightarrow}    
\def\bar{\overline}    
    
\def\be{\begin{equation}}  
\def\ee{\end{equation}}  
\def\bea{\begin{eqnarray}}  
\def\eea{\end{eqnarray}}         
\begin{document}
\begin{flushright}      
Lund-MPh-01/06 
\end{flushright}          
\vspace{0.3 cm}    
\centerline{\Large\bf Probing {\it CPT} violation with atmospheric neutrinos}
\vskip 1 cm 
\begin{center} 
\bf Solveig Skadhauge 
\footnote{Email address: Solveig.Skadhauge@matfys.lth.se}
\end{center} 
\centerline{ \it{Department of Mathematical Physics, LTH, 
Lund University, S-22100 Lund, Sweden}} 
\vskip 1 cm          
\centerline{\bf ABSTRACT}\par       
\vskip 0.5 cm   
We investigate the recently suggested scheme of independent 
mass matrices for neutrinos and antineutrinos. 
Such a {\it CPT} violating scheme is able to account for all neutrino 
data with the three known flavors. 
For atmospheric neutrinos this means that it is possible to 
have different mass squared differences driving the oscillation 
for neutrinos and antineutrinos. We analyze the atmospheric and K2K 
data within the simplest scheme of two neutrino oscillation, 
neglecting electron neutrino oscillation.  
We find that the preferred region is close to the {\it CPT} conserving 
mass spectra. However the spectra with the antineutrino mass squared 
difference about or larger than 0.1 eV$^2$ and the neutrino 
mass squared difference 
about $2 \times 10^{-3}$ eV$^2$ is not significantly disfavored. 
In this parameter region the atmospheric data are 
independent of the antineutrino mass squared difference. 
Therefore no useful constraint can be put on {\it CPT} violation 
effects contributing to different masses for the neutrinos 
and antineutrinos. \\

{\it CPT} violation.
\vskip 1 cm 
\section{Introduction}
Many elementary particles, like the electron and the kaons, provide  
tight bounds on possible {\it CPT} violating effects contributing to 
different masses for the particle and its antiparticle. 
For instance for the electron 
and the positron we have \cite{pgd}
\begin{equation}
  \frac{|m_{\rm e^+} - m_{\rm e^-}|}{m_{\rm average}} < 
  8 \times 10^{-9} \;, \qquad {\rm CL}=90\% \;.
\end{equation}
As is well known, {\it CPT} conservation implies the equality of the 
neutrino and antineutrino survival probabilities in vacuum \cite{petcov}, 
though matter effects can produce fake {\it CPT} violating effects 
\cite{langacker}. 
The atmospheric neutrino data involve both the particle and 
the antiparticle channels and are therefore suitable for a study 
of possible {\it CPT} violation in the neutrino sector. 
The idea to use neutrino oscillation to search for {\it CPT} violation 
was first proposed in Ref.\cite{bigi}.
Naturally as the atmospheric neutrino experiments are probing 
mass squared differences and not the absolute neutrino mass, 
these will be the quantities which might be restricted by the data. 
The interest in {\it CPT} violation arises due to a recently 
suggested scheme which is capable of solving all neutrino anomalies 
without the use of a light sterile neutrino \cite{yan,gabriela}.

At present three neutrino anomalies (atmospheric \cite{SKevi}, 
solar \cite{SKsolar}  and LSND \cite{LSND}) exist, all requiring 
different $\dmsq$'s when interpreted in terms of 
neutrino oscillation. Therefore a {\it CPT} conserving three 
neutrino framework cannot account for all anomalies. 
This has also been explicitely shown in theoretical fits of the atmospheric 
data \cite{nosb,sol2}. Consequently one has to go beyond 
standard explanations to solve all anomalies. 

A possible solution could be the existence of a light sterile neutrino. 
Several studies of such four neutrino models have been performed and the 
current situation has been presented in Ref.\cite{schwetz}. The 
four neutrino models give an acceptable fit when fitting  
all available data. However, each of the different solutions 
faces problems within a particular subset of the data. 
The '3+1' mass spectra are in disagreement with the short-baseline 
experiments and the '2+2' mass spectra conflict with either the 
atmospheric or the solar neutrino data. Therefore the four neutrino models,
though not completely ruled out at present, seem highly disfavored. 

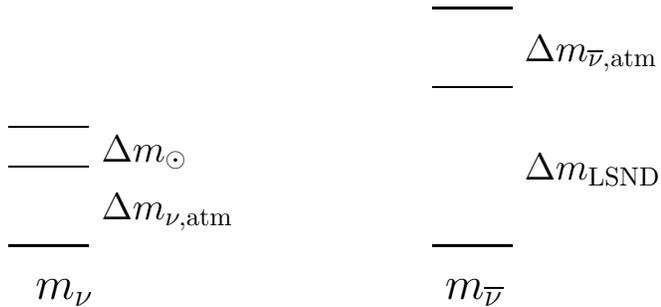
\begin{figure}\label{masses}
\begin{picture}(300,120)(-100,0)
\put(40,20){\line(40,0){30}}
\put(40,50){\line(40,0){30}}
\put(40,65){\line(40,0){30}}
\put(200,20){\line(130,0){30}}
\put(200,80){\line(130,0){30}}
\put(200,110){\line(130,0){30}}
\put(50,0){\Large{$m_{\nu}$}}
\put(205,0){\Large{$m_{\overline\nu}$}} 
\put(75,52){\large{$\D m_{\odot}$}} 
\put(75,30){\large{$\D m_{\nu,\rm atm}$}}
\put(235,45){\large{$\D m_{\rm LSND}$}}
\put(235,90){\large{$\D m_{\bar\nu, \rm atm}$}}  
\end{picture}
\caption{Schematic view of the masses of neutrinos and antineutrinos}
\end{figure}

In the absence of a sterile neutrino, Yanagida and Murayama 
have recently suggested a another possibility to solve all of the known 
neutrino anomalies \cite{yan}. 
The Yanagida-Murayama scheme preserves Lorentz invariance\footnote{It has been 
argued that the scheme also violate Lorentz invariance \cite{lor}} but 
involve {\it CPT} violation by invoking independent masses 
for neutrinos and antineutrinos. 
Hence there is 
a total of four independent $\dmsq$'s. 
Schematically we can represent the masses for the neutrino and 
antineutrino as in Fig.\ref{masses}.  
The solar neutrino problem only concerns the disappearance of 
$\nu_e$ and the LSND experiment sees $\bar \nu_e$ appearance in a 
$\bar\nu_\mu$ beam. These experiments can be separately explained by 
$\dmsq_\odot$ and $\dmsq_{\rm LSND}$ in the Yanagida-Murayama scheme. 
The atmospheric neutrinos data involve both $\nu_\mu$ and 
$\bar\nu_\mu$ and allowing for {\it CPT} violation the two $\dmsq$'s are 
no longer constrained to be identical as also considered in 
Refs.\cite{pakvasa,gabriela}. It is therefore clear that all the data can 
be explained within this model.

To invoke {\it CPT} violation is indeed a very drastic solution. 
Therefore it is important to discuss physical models for generating 
{\it CPT} violation in the Yanagida-Murayama scheme. 
In Ref.\cite{gabriela} a definite model 
of {\it CPT} violation is introduced that can account for  all available 
neutrino data. It was argued that {\it CPT} violation in the 
neutrino sector can be motivated from string theory via the 
extra dimensions. The right-handed neutrinos, like the 
graviton, are free to propagate in the bulk, whereas the Standard Model 
fields are constrained within a four dimensional brane. 
This gives rise to non-locality for the neutrinos and thereby 
generating {\it CPT} violation.  
This {\it CPT} violating scheme is also able to account for 
baryogenesis in a natural way \cite{gabriela}. 
Furthermore non-commutative 
geometry can generate {\it CPT} violation \cite{noncom,yan}.

In this paper we will study the atmospheric neutrino anomaly 
within a two family neutrino scheme with {\it CPT} violation. 
The electron neutrinos are assumed not to oscillate on the 
atmospheric scale. We include the data from the K2K long baseline 
experiment, that further constrain the neutrino parameters. 
The atmospheric neutrino problem is by now well established and can be  
accounted for primarily by a two neutrino $\nu_{\mu}\Ar \nu_{\tau}$ 
oscillation \cite{SKevi}. However, sub-dominant oscillations are 
still possible and maybe even welcome \cite{subdominant,sol2}. 
Having different mass matrices for neutrinos and antineutrinos 
naturally gives different mixing matrices; $U_\nu$ for the neutrino 
sector and $U_{\bar\nu}$ for the antineutrino sector. 
We will investigate whether the mixing parameters can be constrained 
by the atmospheric and K2K data.
A most relevant parameter is the difference 
in mass squared difference for neutrinos and antineutrinos. 
Let us define the parameter 
$\epsilon$ to describe the amount of {\it CPT} violation  
\begin{equation}\label{epsilon}
   \epsilon = | \dmsq_{\nu ,{\rm atm}} - \dmsq_{\bar\nu,{\rm atm}}| \;.
\end{equation}
Using the latest data, we will show that $\epsilon$ is only weakly 
constrained. 

Let us finally mention that the LSND result, which has not yet been confirmed, 
will be scrutinized by the Mini-BooNE experiment \cite{Mini}.  
However, as has been noted before, unless this experiment is done also with
antineutrinos the Yanagida-Murayama scheme cannot be ruled out.
\section{Analysis of the atmospheric data}
  A number of experiments have measured the atmospheric neutrino fluxes. 
Here we will only consider the contained events of the 
Super-Kamiokande (SK) experiment \cite{SKdata}. 
The justification for leaving out other 
data sets is the superior statistics of the SK data.
Furthermore, the high energy upward through-going muon events \cite{SKup} 
are less affected by antineutrinos. For the average energy of 100 GeV of 
these events the $\nu_{\mu}/\bar\nu_\mu$ flux ratio is about 1.5, 
thus decreasing the influence of the antineutrinos. Also the 
statistics is lower and we do not expect large effects from the inclusion 
of this sample. 

We use the following simple two-family survival probability
relations for neutrinos and antineutrinos 
\begin{equation}
  P_{\nu_\mu\rightarrow \nu_\mu}=1-
  \sin^2(2\theta_{\nu})\sin^2 \left(\frac{L\dmsq_\nu}{4E}\right) \;,
\end{equation}

\begin{equation}
  P_{\bar\nu_{\mu}\rightarrow \bar\nu_{\mu}}= 1 - 
  \sin^2(2\theta_{\bar\nu})\sin^2 \left(\frac{L\dmsq_{\bar\nu}}{4E} \right)\;.
\end{equation}
We assume that the oscillation is into $\tau$-neutrinos, whereby 
the electron survival probability is taken to be one for both neutrinos and 
antineutrinos. As we only consider $\nu_\mu$ to $\nu_\tau$ 
oscillation there are no matter effects. 
The pathlength of the neutrino, $L$, is calculated using an 
average production point in the atmosphere of 15km. $E$ is the neutrino 
energy.

The data are divided into sub-GeV and multi-GeV energy ranges 
and can be represented as the ratio, $R^{\rm exp}$, 
between the measured fluxes and the theoretical Monte Carlo 
prediction in the case of no oscillation. 
We define $\chi^2$ as
\be
  \chi^2= \sum_{M, S}\sum_{\alpha=e,\mu}\sum_{i=1}^{10} 
     \frac{(R_{\alpha,i}^{\rm exp}-
     R_{\alpha,i}^{\rm th})^2}{\sigma_{\alpha i}^2} + \chi^2_\beta \;,
\ee
where $\sigma_{\alpha,i}$ are the statistical errors 
and $M,S$ stand for the multi-GeV and sub-GeV data respectively and $i$ 
denotes the zenith angle bin. 
For the details of the $\chi^2$ definition we refer to Ref.\cite{sol2}, 
except that we here use a smearing of the sub-GeV events with an angle 
$50^\circ /\sqrt{E_\nu/{\rm GeV}}$ \cite{learned}. 
The overall normalization of the neutrino fluxes is allowed to vary freely. 
Hence we minimize with respect to $\alpha$, where the neutrino flux is 
given by $\Phi=(1+\alpha )\Phi^0$. The theoretically predicted neutrino 
flux $\Phi^0$ is taken from \cite{subflux,mulflux}. 
The $\chi^2_\beta$ term takes into account the error in 
the $\nu_\mu/\nu_e$ flux ratio. 
The SK Collaboration estimate the error to be $8\%$ in the sub-GeV 
range and $12\%$ in the multi-GeV range. We renormalize the neutrino 
fluxes as 
\begin{equation}
  \tilde\Phi_{\mu}^{S,M}=(1-\b_{S,M}/2)\Phi_{\mu}^{S,M}  \; , \qquad
  \tilde\Phi_{e}^{S,M}=(1+\b_{S,M}/2)\Phi_{e}^{S,M}      
\end{equation}
where the symbols $S,M$ stands for sub-GeV and multi-GeV respectively 
and minimize the total $\chi^2$ function with respect to $\b_S$ and $\b_M$. 
The $\chi^2_\beta$ function is given by
\begin{equation}
  \chi^2_{\beta}=\left( \frac{\beta_S}{0.08}\right) ^2
               +\left( \frac{\beta_M}{0.12}\right) ^2 \;.
\end{equation}
For the best fit point obtained by the SK Collaboration the values used 
are $\beta_S=6\%$ and $\beta_M=12\%$, implying that while scaling 
the electron ratios down one simultaneously scales the muon ratios up. 

We also include the recent data from the K2K long baseline 
experiment \cite{k2k}.
The beam is almost pure $\nu_\mu$ and we will neglect the small contamination 
of $\bar\nu_\mu$ and $\nu_e$. We use the same method as in 
Ref.\cite{foglik2k} and only fit to total number of observed events. 
In total we have five parameters ($\dmsq_{\nu},\; \dmsq_{\bar\nu},\; 
\theta_\nu,\; \theta_{\bar\nu},\;\alpha$) and 41 data points.

\begin{figure}[t]
\begin{center}
\mbox{
\epsfysize=9cm
\epsffile{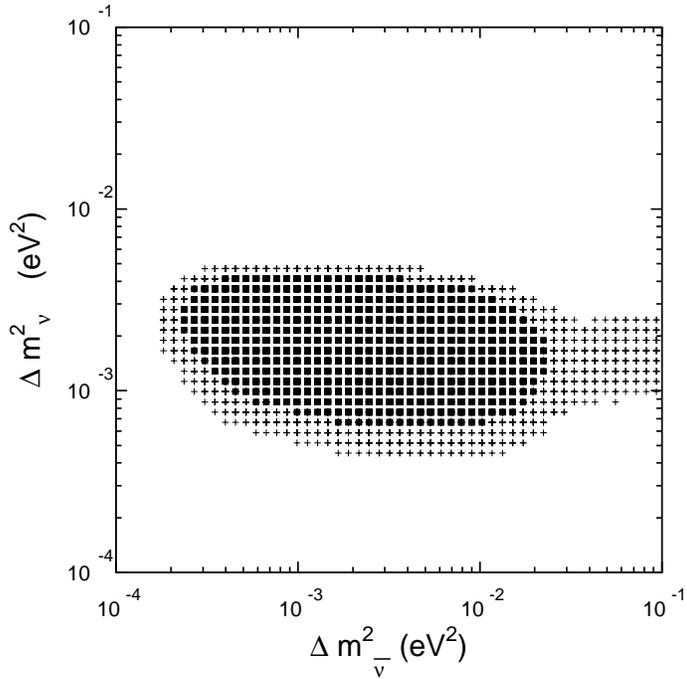}
}
\end{center}
\vspace{-0.4cm}
\caption{The 68.3\% (90\%) CL regions for parameters $\dmsq_{\nu}$ and 
$\dmsq_{\bar\nu}$  for the contained SK events and the total number 
of events in K2K (5 d.o.f). 
}
\label{conlev}
\end{figure}
The minimum is $\chi^2_{\rm min}=33$ at $\alpha=1\%$, 
$\beta_S=8\%$, $\beta_M=10\%$ and  
\begin{equation}\label{bcpt}
 \dmsq_{\nu}=2.5 \times 10^{-3} \:{\rm eV}^2 \;, \;\; 
 \dmsq_{\bar \nu} = 2.0 \times 10^{-3} \:{\rm eV}^2 \; , \;\;
 \sin^2(2\theta_{\nu})=1.0 \; , \;\;
 \sin^2(2\theta_{\bar\nu})=1.0 
\end{equation}
with 36 degrees of freedom. 
This is very close to the {\it CPT} conserving case. 
In Fig.\ref{conlev} we show the 68.27\% and 90\% confidence levels 
as obtained by $\Delta \chi^2 <5.9,\; 9.2$, respectively, 
for five degrees of freedom. At 90\% C.L.  
the mass squared difference for neutrinos is 
constrained within $4.5 \times 10^{-4}$ eV$^2-5 \times 10^{-3}$ eV$^2$, 
while the antineutrino mass squared difference is only bounded from below 
($>2\times 10^{-4}$ eV$^2$).
At 90\% C.L. the mixing angles are bounded, $\sin^2(2\theta_\nu)>0.8$ and 
$\sin^2(2\theta_{\bar\nu})>0.5$, but maximal mixing is preferred for both 
angles. 

  The correlation between the lepton and the neutrino angle in the sub-GeV 
range is very weak and the data is smeared compared to the multi-GeV 
sample \cite{review}. The exact calculation method 
can therefore change the results slightly as the effect of a 
large $\dmsq_{\bar\nu}$ is similar to a smearing. For smaller effective 
smearing the best fit point will move toward larger values 
of $\dmsq_{\bar\nu}$. The same considerations apply to the mixing angles 
which also effectively flatten the zenith angle curve. 

One should note that a point with values of $\dmsq_{\bar\nu} \simeq$ 
0.1 eV$^2$ and $\dmsq_\nu \simeq 2 \times 10^{-3}$ is not significantly 
disfavored. The SK contained data become independent of $\dmsq_{\bar\nu}$ in 
this region as the oscillation probabilities are averaged to 1/2 for all 
pathlengths. K2K is obviously independent of $\dmsq_{\bar\nu}$ as they measure 
neutrinos. 
\begin{figure}[t]
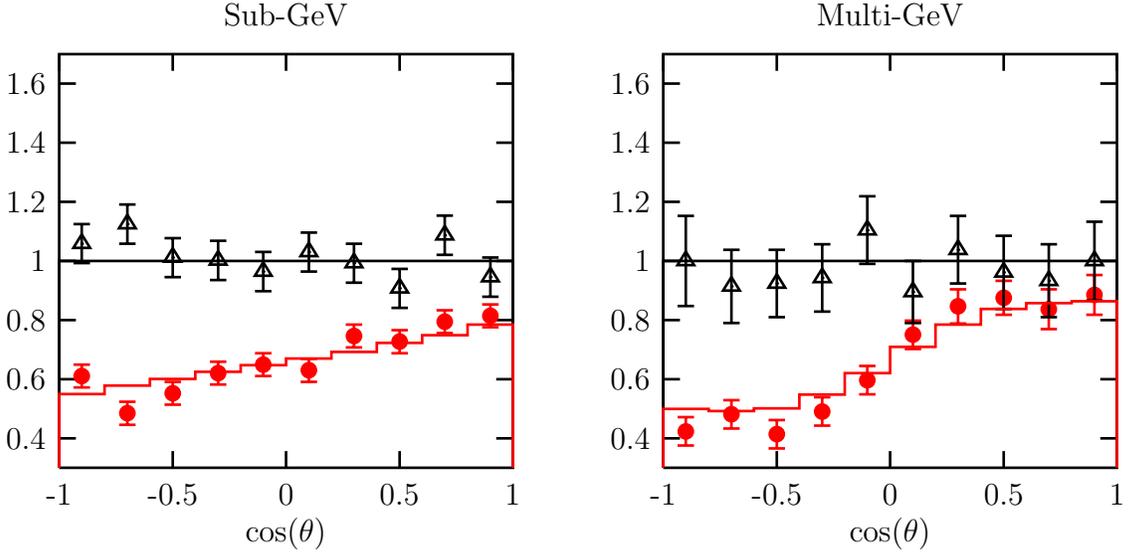

\mbox{
\input{ratio_s.ps}
}
\mbox{
\input{ratio_m.ps}
}
\caption{Predicted ratios as a function of the zenith angle for mixing 
parameters; $\dmsq_{\nu}=2\times 10^{-2}$ eV$^2$, 
$\dmsq_{\bar\nu}=0.1$ eV$^2$ and maximal mixing, and   
the data with statistical errors. The upper 
curves are the $\nu_e$ ratios and the lower curves are the $\nu_\mu$ ratios. 
The triangles are the $\nu_e$ experimental ratios and the circles are the 
$\nu_\mu$ experimental ratios. Note that the experimental ratios are 
plotted using $\tilde \Phi$ and not $\Phi^0$.
} 
\label{ratio}
\end{figure}
The predicted ratios of this mass spectra are shown in 
Fig.\ref{ratio} along with the data points. The predicted 
ratio of around 0.85 for the downward going multi-GeV 
muon neutrinos can be easily understood. 
In this energy range the flux of neutrinos is roughly the same as 
the flux of antineutrinos. But the antineutrino cross section is 
less than half that of neutrinos. The $\sin^2(L\dmsq/4E)$ 
is averaged to one half for antineutrinos and to one for neutrinos  
and the ratio is estimated to
\begin{equation}
   R_{\mu,\downarrow} 
          \simeq \frac{\Phi_{\nu_\mu}P_{\nu_\mu \Ar \nu_\mu} \sigma_{\nu_\mu}
          +\Phi_{\bar\nu_\mu} P_{\bar\nu_\mu \Ar \bar\nu_\mu} \sigma_{\bar\nu_\mu}}
          {\Phi_{\nu_\mu} \sigma_{\nu_\mu} + \Phi_{\bar\nu_\mu}\sigma_{\bar\nu_\mu}}
           \simeq 0.85\;, 
\end{equation}
where $\sigma$ is the cross section. The reason that this mass spectra 
is not strongly disfavored is because it agrees very well with the well 
known double ratio. The measured value is  
\begin{equation}\label{double}
R=\frac{(\mu/e)_{\rm DATA}}{(\mu/e)_{\rm MC}} = 0.675^{+0.034}_{-0.032}\pm0.080
\end{equation}
in the multi-GeV range \cite{SKdata}. 
The prediction double ratio for $\dmsq_\nu \simeq 2 \times 10^{-3}$ eV$^2$ 
and $\dmsq \simeq 0.1$ eV$^2$ are $R \simeq 0.68$, whereas for both 
$\dmsq$'s around $3 \times 10^{-3}$ eV$^2$ we get $R \simeq 0.75$. 
In both cases we have assumed maximal mixing. 
For sub-GeV events the double ratio is $0.638\pm 0.017 \pm 0.050$.  
In other word the advantage of the {\it CPT} violating mass spectra is that 
it for a flux normalization that diminishes the excess of in particular 
sub-GeV $\nu_e$ events also agrees well with the muon ratios.  
In fact in the case that the overall normalization is 
varied freely, but the $\mu/e$ ratio is kept fixed, the best fit point 
is for $\dmsq_{\bar\nu} \gg \dmsq_\nu$. Although such a mass spectra 
does not fit the up-down asymmetries that well as seen from Fig.\ref{ratio}. 
It must be remembered that the measured values quoted in 
Eq.(\ref{double}) is dependent on the theoretical predictions for 
the $\nu_\mu/\nu_e$ flux ratio. The double ratio basically describes 
the average gap between the $\nu_e$ and the $\nu_\mu$ ratios and 
this gap is somewhat too large to be fitted very well by a {\it CPT} 
conserving two family $\nu_{\mu}\Ar \nu_\tau$ oscillation. 
However if the theoretically predicted $\nu_\mu/\nu_e$ flux 
ratio is decreased by 6-12\% this provides a very good fit to the data. 
Furthermore the low value of the double ratio could be due to 
an excess of $\nu_e$ events which is not accounted for by the two 
family scheme.
The $\chi^2$ value for the point in Fig.\ref{ratio} is 39 and therefore 
barely outside the $1\sigma$ region.  When using 2 d.o.f. as in 
Ref.\cite{strumia} this would be a $2\sigma$ exclusion. We remark that there 
is a local maximum of the $\chi^2$ function for values of $\dmsq_{\bar\nu}$ 
between $10^{-2}-10^{-1}$ eV$^2$. The main differences between our results and 
those in Ref.\cite{strumia} are due to the fact that we use the 
theoretical predicted fluxes, whereas the normalization in each type of 
events are varied freely in Ref.\cite{strumia}. 

For almost all mixing parameters a rise of the theoretical fluxes are 
needed. New calculations suggest instead a lower flux, 
by about 10\% \cite{newflux}, mainly due to the primary flux being lower 
than obtained in earlier measurements. 
The new normalization results in a large excess of $\nu_e$ events 
which seems to be difficult to obtain theoretically. In the conventional 
{\it CPT} conserving case the SK Collaboration obtains $\alpha=20\%$ 
for the best fit point using the new flux predictions \cite{lipari}, 
which roughly amount to putting the normalization back to the old value. 
Therefore we find the error of the overall normalization is large 
and a rise can not be excluded. 

In the {\it CPT} violating scheme the electron ratios can be considerably 
away from one. A relatively large LSND angle \cite{sol2} as well as 
the solar mass squared difference \cite{smirnov} can influence these ratios. 
Though the LSND angle is constrained to be small by the BUGEY results. 
Also the two angles $\theta_{e\mu}^\nu$ and $\theta_{e\mu}^{\bar\nu}$, 
describing the oscillation of $\nu_e$ driven by the atmospheric mass squared 
difference, can have effects. 
In particular $\theta_{e\mu}^\nu$ is not constrained by the CHOOZ 
\cite{chooz} and Palo Verde \cite{palo} results as these experiments are 
measuring anti neutrinos. Moreover $\theta_{e\mu}^{\bar\nu}$ can be large if 
$\dmsq_{\bar\nu}$ is below the CHOOZ sensitivity of $10^{-3}$ eV$^2$, which 
is not ruled out by the present data (see Fig.\ref{conlev}).
The influence of some of these extra mixing parameters has been studied in 
Ref.\cite{gabriela2}, where however the systematic errors in the SK data 
have been ignored.  

As we have shown the limits on {\it CPT} violation in the neutrino sector 
are rather weak at present. There are nevertheless good prospects for a much 
stronger bound in the near future. The results from the KamLAND experiment 
\cite{kamland} could likely disprove the Yanagida-Murayama scheme. The 
experiment will test the currently favored large mixing angle MSW (LMA) 
solution to the solar neutrino problem, 
by detecting anti electron neutrinos from nearby nuclear reactors.  
For the most favored region of the LMA solution 
($\dmsq_{\odot} < 2 \times 10^{-4}$ eV$^2$) 
the detected energy spectrum at KamLAND will quite precisely determine 
the value of the mass squared difference and this signal would rule out the 
Yanagida-Murayama scheme. However for $\dmsq > 2 \times 10^{-4}$ eV$^2$ 
the oscillations are averaged out and KamLAND can only put a lower bound 
on the anti neutrino mass squared difference \cite{barger}. 
In this case the situation becomes more problematic as there are two different 
possible explanations. 
A large value of $\dmsq_{\odot}$ is not ruled out, though disfavored by the 
present data. 
The signal could also be explained within the Yanagida-Murayama scheme by 
having a small $\dmsq_{\bar\nu, atm}$ along with a large 
$\theta_{e\mu}^{\bar\nu}$. Borexino detecting solar neutrinos will not be 
able to pin down the true solution. 
Hence, if an averaged oscillation with a large angle 
is observed, one would most likely have to wait for the results of the 
MiniBooNE experiment. In the case that KamLAND does confirm the LMA 
solution to the solar neutrino problem, much better 
limits on {\it CPT} violation could be obtained as discussed 
in Ref.\cite{bahcall}. If Kamland does not observe a suppression of the anti 
neutrino flux there are different possibilities to test {\it CPT} violation 
as discussed in Refs.\cite{strumia,gabriela2}. 
Indirect limits can also be obtained by studying how radiative corrections 
communicate the large amount of {\it CPT} violation in the neutrino sector 
to the charged lepton sector \cite{irina}.

  In conclusion we have analyzed the Super-Kamiokande contained events 
and the K2K data in a {\it CPT} violating two neutrino 
$\nu_{\mu},\,\nu_{\tau}$ framework. 
The best fit area is close to the {\it CPT} conserving case. 
However at present the {\it CPT} violation parameter $\epsilon$, 
defined in Eq.(\ref{epsilon}), cannot be usefully constrained. 

\vskip 1.0cm
{\bf Acknowledgment}\\ 
The author expresses her thanks to Gabriela Barenboim and Amol Dighe for 
collaboration when developing the program used in this analysis. 
Furthermore the author is grateful to Cecilia Jarlskog for discussions 
and encouragements. 
\vskip 0.5cm     
  
\end{document}